\documentclass[12pt]{article}
\usepackage{amsmath,amsfonts,amssymb,latexsym}
\usepackage{graphicx,multirow}
\usepackage{subfig}
\begin{document}
\begin{center}

{ \bf FINITE TEMPERATURE CASIMIR EFFECT IN RANDALL-SUNDRUM MODELS}

\bigskip

Marianne Rypest{\o}l and Iver
Brevik\footnote{iver.h.brevik@ntnu.no}

\bigskip

Department of Energy and Process Engineering, Norwegian University
of Science and Technology, N-7491 Trondheim, Norway

\bigskip


\begin{abstract}

The finite temperature Casimir effect for a scalar field in the
bulk region of the two Randall-Sundrum models, RSI and RSII, is
studied. We calculate the Casimir energy and the Casimir force for
two parallel plates with separation $a$ on the visible brane in
the RSI model. High-temperature and low-temperature cases are
covered. Attractiveness versus repulsiveness of the temperature
correction to the force is discussed in the typical special cases
of Dirichlet-Dirichlet, Neumann-Neumann, and Dirichlet-Neumann
boundary conditions at low temperature. The Abel-Plana summation
formula is made use of, as this turns out to be most convenient.
Some comments are made on the related contemporary literature.

\end{abstract}

\end{center}


\section{Introduction}

Inspired by the Randall-Sundrum models \cite{randall99}, there has
recently been  considerable interest in the Casimir effect in
higher-dimensional space. We may recall the characteristic
features of this model: in the first variant, called RSI, one
assumes that we are living on a (3+1)-dimensional subspace called
a 3-brane, separated from an additional hidden brane by a bulk
region. Only gravity is assumed to propagate in the bulk. The
extra dimension is a circle $S^1$ with radius $r_c$, represented
by a coordinate $\phi$ in the range $-\pi \leq \phi \leq \pi$. The
hidden and the visible branes are located at $\phi=0$ and
$\phi=\pi$ respectively. Imposition of $Z_2$ symmetry means that
the points $(x^\mu,\phi)$ and $(x^\mu, -\phi)$ are identified. In
the second variant of the model, RSII, the hidden brane is sent to
infinity. The major difference between the RS model and other
higher-dimensional models lies in the warp factor
$e^{-2kr_c|\phi|}$ in the metric
\begin{equation}
ds^2=e^{-2kr_c|\phi|}\eta_{\mu\nu}dx^\mu dx^\nu-r_c^2 d\phi^2,
\label{1}
\end{equation}
where $\eta_{\mu\nu}=$diag(1,-1,-1,-1) is the Minkowski metric of
flat spacetime and $k$ is a constant of order $M_{Pl}$, the Planck
mass. The warp factor plays an important role helping to solve the
hierarchy problem without introducing additional hierarchies.

Whereas in the original RS model the fields of the standard model
(SM) as mentioned were confined to the visible brane only, the
possibility of having additional fields in the bulk was soon
investigated, beginning with scalar fields
\cite{goldberger99,flachi01} introduced to stabilize the
inter brane distance. Subsequently, the possibility of having other
fields, such as fermion field \cite{grossman00,gherghetta00},
gauge fields \cite{pomarol00,davoudiasl00,gherghetta00}, or even
the full assembly of SM fields \cite{davoudiasl01}, was
investigated. Tests of Newton's law at short distances may show
aberrations at short distances (cf., for instance,
Refs.~\cite{callin05}).

Only recently have there appeared papers on the Casimir effect in
the Randall-Sundrum models. To our knowledge the first group
working on this was Frank {\it et al.}, publishing two papers
\cite{turan07,saad08} on the Casimir force in both the RSI/RSII
and the RSI-q/RSII-q models. Here RSI-q and RSII-q refer to
generalizing the 3-branes on RSI/RSII to $(3+q)$-branes (this kind
of generalization will however not be dealt with  in the present
paper). There are also two papers of the group of
Morales-T\'{e}cotl {\it et al.} \cite{morales08}, focusing on
RSI-q/RSII-q. While Frank {\it et al.} used zeta function
regularization to calculate the Casimir force, Morales-T\'{e}cotl
{\it et al.} used a Green's function formalism. A delicate point
is that their results are seemingly in conflict except in the
Minkowskian case. We shall comment on this point later.

A third class of papers  are those of Cheng
\cite{cheng09,cheng09a,cheng09b}. These papers, except from
Ref.~\cite{cheng09b}, assume zero temperature. Moreover, Elizalde
{\it et al.} have recently studied repulsive Casimir effects from
extra dimensions for a massive scalar field with general curvature
parameter \cite{elizalde09c}.

Among previous papers in this research field, the ones most
similar in nature to that of the present work  are those of Teo
\cite{teo09m,teo09n,teo09,teo09a,teo09b}. She calculated the
temperature Casimir force; not, however, the Casimir free energy.

As already mentioned we will assume a scalar field $\Phi$ in the
bulk. Both the RSI and the RSII models will be considered.
Formally, the  expressions pertaining to  the RSII case may be
derived  by letting $r_c \rightarrow \infty$ in the RSI
expressions. The use of a scalar field makes of course the
situation more unphysical that what would be the case by assuming
an electromagnetic field in the bulk. But we avoid the
complications arising from photon spin in higher-dimensional
spacetimes (for some recent papers in that direction see, for
instance,
Refs.~\cite{alnes07,brevik08,hofmann08,edery08,pascoal08,poppenhaeger04}).
The conflict is not even resolved when we have only one extra
spatial dimension and spacetime is {\it flat}. On one hand we have
e.g. Poppenhaeger {\it et al.} \cite{poppenhaeger04} and Pascoal
{\it et al.} \cite{pascoal08}. They find the electromagnetic
Casimir force by multiplying the scalar field expressions by a
factor $p$ (to account for the possible polarisations of the
photon) and subtract the mode polarised in the direction of the
brane. On the other hand we have Edery and Marachevsky
\cite{edery08} who start out with decomposition of the
five-dimensional Maxwell action. This conflict in not the central
issue of this paper, and we avoid it by consider scalar fields
only.

Our main purpose
will be to calculate the Casimir free energy and the Casimir force
for the RSI model, when there are two parallel plates with
separation $a$ on the visible brane. This is the piston model.
 Our main focus will be on the following
points:

\bigskip

1) The calculation is given for arbitrary temperature $T$, and the
low-temperature and high-temperature limits are thereafter
considered.  The attractiveness versus the repulsiveness of the
temperature corrections for different boundary corrections at low
temperature are of definite physical interest and are therefore
pointed out. We regularize infinite expressions by using zeta
functions and the Abel-Plana summation formula, as this formula
turns out to be better suited to the problem than the more
commonly used Euler-Maclaurin formula.

2)  We assume Robin boundary conditions on the physical plates at
$x=0$ and $x=a$. Usually, one has been considering the more simple
Dirichlet  conditions when working on this kind of problems,
although very recently  the Robin considerations have begun to
attract attention   \cite{elizalde09,teo09b}.

\bigskip

As an introductory step, we consider in the next section the
partition function and the free energy of a bulk scalar field. We
discuss the distinction between even and odd fields, and consider
also the mode localization problem. After a brief survey of the
Abel-Plana formula in Section 3 we consider in Section 4 the
Dirichlet-Dirichlet (DD), Neumann-Neumann (NN), and
Dirichlet-Neumann (DN) boundary conditions in flat space, at finite
temperature without extra dimensions.
Our main topic, the temperature RSI case, is covered
in Section 5, where the Casimir free energy and force are
calculated for the different boundary conditions.  A brief
treatment of the RSII case is given in Section 6.

It should be recognized that the warp factor is an important
element in the present problem. One might analyze instead the
analogous higher-dimensional cases taking spacetime to be {\it
flat}. Considerable interest has been devoted to this simpler
variant of higher-dimensional Casimir theories in recent years.
See, for instance,
Refs.~\cite{teo09,fulling09,edery08,edery08a,fulling09a,rypestol09,pascoal08},
and further references therein.

\section{Free energy of a bulk scalar field}

To find the partition function for a non-minimally coupled scalar
field $\Phi$ with mass $m$ in the RSI model, we follow a
Kaluza-Klein reduction approach \cite{ichinose09},
starting from the Lagrangian density
\begin{equation}
{\cal L}= \sqrt{-G} \left( \frac{1}{2} \partial_M \Phi \partial^M \Phi
-\frac{1}{2} (m^2+\zeta R+c_\mathrm{hid}\delta(z)
+ c_\mathrm{vis}\delta(z-z_r)) \Phi^2 \right).  \label{2}
\end{equation}
Here $G$=det $G_{MN}$ (with $M,N=0,1,2,3,5$) is the determinant of the 5D metric,
$R$ is the 5D Ricci scalar, $\zeta$ is the conformal coupling, and
$c_{hid/vis}$ are the boundary mass terms on the branes.
Throughout the article we use $\hbar=c=k_B=1$. We have
introduced above  a new position coordinate $z$ such that
$|z|=(e^{k|r_c\phi|}-1)/k$, implying that $z_r=(e^{k\pi
r_c}-1)/k$.  It is convenient to introduce also
 the quantity $A(z)=1/(1+k|z|)$. The partition function
\begin{equation}
Z = \int \mathcal{D} \Phi \exp \left( i \int \mathrm{d}^4x \mathrm{d}z \mathcal{L} \right) \label{3}
\end{equation}
can now be calculated, making use of the Euclideanization
$\tilde{x}^i=x^i, \tilde{x}^0=\tau=ix^0$. A partial integration
yields
\begin{equation}
Z=\int {\cal{D}} \Phi \exp\left[ -\int
d^4\tilde{x}dz\frac{1}{2}\Phi A^3(z)\left(
\hat{p}^2+\hat{M}_z^2\right) \Phi\right]. \label{4}
\end{equation}
Here
\begin{equation}
\hat{p}^2=\tilde{\eta}^{\mu\nu}\partial_\mu\partial_\nu, \label{5}
\end{equation}
where $\tilde{\eta}^{\mu\nu} =-\delta^{\mu\nu}$ is the metric in
the coordinates $\tilde{x}^\mu$, and
\begin{equation}
\hat{M}_z^2=A^{-3}(z)\left[
-\partial_zA^3(z)\partial_z+A^5(z)\left(m^2+\zeta
R+c_{hid}\delta(z)+c_{vis}\delta(z-z_r)\right) \right]. \label{6}
\end{equation}
We now expand $\Phi$ in the eigenfunctions $\chi_p(\tilde{x}^\mu)$
and $\psi_N(z)$ of $\hat{p}^2$ and $\hat{M}_z^2$ respectively,
\begin{equation}
\Phi(\tilde{x},z)=\sum_{N,p} c_N(p)\chi_p(\tilde{x})\psi_N(z),
\label{7}
\end{equation}
where $(\tau=it)$
\begin{equation}
\hat{p}^2\chi_p(\tilde{x})=-(\partial_\tau^2+\partial_x^2+\partial_y^2+\partial_z^2)\chi_p(\tilde{x})=p^2\chi_p(\tilde{x}),
\label{8}
\end{equation}
\begin{equation}
\hat{M}_z^2\psi_N(z)=M_N^2\psi_N(z). \label{9}
\end{equation}
The eigenfunctions are normalized as
\begin{equation}
\int
d^4\tilde{x}\,\chi_p(\tilde{x})\chi_{p'}(\tilde{x})=\delta_{pp'},
\label{10}
\end{equation}
\begin{equation}
\int_{-z_r}^{z_r}dz\,\psi_N(z)A^3(z)\psi_{N'}(z)=\delta_{NN'}.
\label{11}
\end{equation}
The partition function now takes the form (an unimportant factor
omitted)
\begin{equation}
Z=\prod_{M_N,p}\left( M_N^2+p^2\right)^{-1/2}, \label{12}
\end{equation}
where the sum goes over all eigenvalues of $M_N$ and $p$. Our next
step is now to identify $M_N$ and $p$.

\subsection{Eigenfunctions and eigenvalues for $\hat{p}^2$ and
$\hat{M}_N^2$ }
\label{Section21}

We start from Eq.~(\ref{8}), assuming Robin boundary conditions on
the physical walls,
\begin{equation}
\left. (1 + \beta_0 \partial_x)\chi_p(\tilde{x}) \right|_{x=0} = 0, \label{13}
\end{equation}
\begin{equation}
\left. (1 - \beta_a \partial_x)\chi_p(\tilde{x}) \right|_{x=a} = 0, \label{14}
\end{equation}
with constants $\beta_0$ and $\beta_a$ referring to $x=0$ and
$x=a$. The forms above are as in Ref.~\cite{elizalde09}. Dirichlet
and Neumann boundary conditions correspond to $\beta=0$ and
$\beta=\infty$, respectively. We assume eigenfunctions of the form
\begin{equation}
\chi_p(\tilde{x}) = N e^{i (\epsilon_l \tau + k_y y + k_z z)} \cos (k_x x +\alpha), \label{15}
\end{equation}
with eigenvalues
\begin{equation}
p^2 = \epsilon_l^2 + k_x^2 + k_y^2 + k_z^2. \label{16}
\end{equation}
For a bosonic field at temperature $T$ the Matsubara frequencies
are
\begin{equation}
\epsilon_l= 2 \pi Tl, \quad l \in \mathbb{Z} \label{17}
\end{equation}
Equation (\ref{13}) leads to the following constraint on $\alpha$,
\begin{equation}
\cos\alpha=\frac{\beta_0k_x}{\sqrt{1+\beta_0^2k_x^2}}, \label{18}
\end{equation}
whereas Eq.~(\ref{14}) yields $F_x(k_x)=0$, where
\begin{equation}
F_x(k_x) =
(1-k_x^2\beta_0\beta_a)\sin(k_xa)-k_x(\beta_0+\beta_a)\cos(k_xa).
\label{19}
\end{equation}
Consider next the eigenfunctions $\psi_N(z)$. We insert the
expression (\ref{6}) into Eq.~(\ref{9}), take into account that
the Ricci scalar for the RS metric is
\begin{equation}
R= -20k^2 +16k( \delta (z) - \delta (z-z_r) ), \label{20}
\end{equation}
and change the position coordinate in the bulk back to $y$ using
$d/dz=A(y)d/dy$. Then,
\begin{equation}
\psi_N''(y) - 4k \psi_N'(y) + \left(M_N^2 e^{2k y} -(m^2-20\zeta k^2) \right) \psi_N(y) =0. \label{21}
\end{equation}
The solution is (we consider the region $0<y<\pi r_c$ only)
\begin{equation}
\psi_N(y) =  \frac{e^{2 k y}}{C_N} \left( J_\nu \left( \frac{M_N}{k} e^{ k y} \right) + b_\nu(M_N) Y_\nu \left( \frac{M_N}{k} e^{k y} \right) \right), \label{22}
\end{equation}
where $\nu=\sqrt{4+(m/k)^2-20\zeta}$ and $C_N$ a normalization constant.
This is the same result as
in Ref.~\cite{gherghetta00}, except that we include curvature
$(\zeta \neq 0)$ in our model.

One should now distinguish between even fields satisfying
$\psi_N(-y)=\psi_N(y)$, and odd fields satisfying
$\psi_N(-y)=-\psi_N(y)$. The behavior may be summarized as
follows:
\begin{itemize}
\item Even scalar fields obey the Robin BC on the branes. If the
field is minimally coupled $(\zeta=0)$ and there is no mass
boundary term $(c_{brane}=0)$, the boundary condition reduces to
the Neumann BC, $\psi_N'(y)|_{brane}=0$. \item Odd scalar fields
obey the Dirichlet BC on the branes.
\end{itemize}
These two cases may be combined: we introduce the two functions
\begin{equation}
\begin{split}
j^\mathrm{brane}_\nu(z) &= (2-(k\beta_\mathrm{brane})^{-1})J_\nu(z)+zJ_\nu'(z) \\
y^\mathrm{brane}_\nu(z) &= (2-(k\beta_\mathrm{brane})^{-1})Y_\nu(z)+zY_\nu'(z),
\label{23}
\end{split}
\end{equation}
and let now $z$  mean $z=e^{k\pi r_c} M_N/k$ (not to be mixed up
with the coordinate $z$ in Sect. 2), and  $d=e^{-k\pi r_c}$. Then
we can write the general BC as $F_N(z)=0$, where
\begin{equation}
F_N(z) = j^\mathrm{hid}_\nu(zd)y^\mathrm{vis}_\nu(z)-j^\mathrm{vis}_\nu(z)y^\mathrm{hid}_\nu(zd). \label{24}
\end{equation}
This is in accordance with Ref.~\cite{brevik01} in the case of
minimal coupling, if we choose $c_{hid}=-c_{vis}=2\alpha/k$.

Special attention ought to be given to the massless case, $M_N=0$.
For fields with $m^2-20\zeta k^2\neq 0$ there is no solution of
Eq.~(\ref{21}) with $M_N=0$ satisfying the Robin BC on both
branes. For an even field with $m^2-20\zeta k^2\neq 0$ with no
boundary mass term the situation is different, as $\psi_0=$const
is a solution of Eq.~(\ref{21}) and also satisfies the boundary
condition which in that case is the Neumann BC. The $M_N=0$ case
has important consequences for the Casimir force from a bulk
scalar field. This is related to the localization problem for the
Kaluza-Klein modes in general. In RSI, the massless mode is
localized near the hidden brane at $y=0$. In RSII the situation is
reversed, as the massless mode is  localized near the visible
brane at $y=0$ and the massive modes are delocalized. The reader
may consult Refs.~\cite{morales08,flachi09} for a discussion as to
what weight to be given to the massless modes in RSI due to the
fact that it is localized near the hidden brane only.

\subsection{Approximate expressions for the masses}

We assume $d=e^{-k\pi r_c} \ll 1$ but keep $z=e^{k\pi r_c}M_N/k$
arbitrary, to find convenient approximative expressions for the
Kaluza-Klein masses. As in this case $j_\nu^{brane}(z) \ll
y_\nu^{brane}(z)$, the equation $F_N(z)=0$  reduces to
\begin{equation}
j_\nu^{vis}(z)=0. \label{25}
\end{equation}
The situation can be divided into two classes:

(i) For {\it Dirichlet} BC $(\beta_{brane}=0)$ it follows that we
need the zeros of $J_2(z)$. Making use of the large-$z$
approximation $J_\nu(z) \sim (2/\pi)^{1/2}\cos
(z-\frac{1}{2}\nu\pi -\frac{1}{4}\pi)$ we find that the expression
\begin{equation}
M_N = k\pi e^{-k\pi r_c} \left( N + \frac{1}{2}\nu - \frac{1}{4} \right), ~N=1,2,\ldots \label{26}
\end{equation}
is useful for practical purposes.

(ii) For {\it non-Dirichlet} BC $(\beta_{brane} \neq 0)$ we obtain
from Eq.~(\ref{23})
\begin{equation}
j_\nu(z) = (2+\nu- (k\beta)^{-1} ) J_\nu(z) - zJ_{\nu+1}(z), \label{27}
\end{equation}
leading approximately to
\begin{equation}
M_N = k\pi e^{-k\pi r_c} \left( N +\frac{1}{2}\nu -\frac{3}{4} \right), ~N=1,2,\ldots \label{28}
\end{equation}
The formula is good for $k\beta_{brane}>1$ and becomes better for higher
$N$. As an example, choosing $k\beta_{brane}=10^3, d=10^{-12},
\nu=2$, we find the numerical error of the zeros to be about 4\% when $N=3$ and
around 1\% when $N=5$. As we will see later the first (i.e. smallest) values
of $M_N$ are the most significant for the Casimir force in RSI.

\subsection{Two expressions for the free energy}

From Eq.~(\ref{12}) we obtain for the free energy
\begin{equation}
F=-T\ln Z=\frac{1}{2}TV_\perp
\int\frac{d^2k_\perp}{(2\pi)^2}\sum_{l=-\infty}^\infty
\sum_{M_N,k_x}\int\ln(M_N^2+\epsilon_l^2+k_x^2+k_\perp^2),
\label{29}
\end{equation}
where $k_\perp^2=k_y^2+k_z^2$, $V_\perp$ is the transverse volume,
$\epsilon_l=2\pi Tl$, and the summations over $k_x$ and $M_N$ go
over all real zeros of the functions $F_x(k_x)$ (Eq.~(\ref{19}))
and $F_N(z)$ (Eq.~(\ref{24})).

By making use of the zeta function
\begin{equation}
\zeta(s) = \sum_{l=-\infty}^\infty V_\perp \int
\frac{\mathrm{d}^2k_\perp}{(2\pi)^2}  \sum_{M_N,k_x} \left( M_N^2
+ \epsilon_l^2 + k_x^2 + k_\perp^2 \right)^{-s}, \label{30}
\end{equation}
following \cite{geyer08}, we can re express $F$ as
\begin{equation}
F=-\frac{1}{2}T\frac{\partial}{\partial
s}\mu^{2s}\zeta(s)\Big|_{s=0}, \label{31}
\end{equation}
where $\mu$ is an arbitrary parameter with dimension mass.

We now derive the classical expression for $F$ using that the
Mellin transform of $b^{-z}\Gamma(z)$ is $e^{-bt}$, i.e.,
\begin{equation}
b^{-z} = \frac{1}{\Gamma (z)} \int_0^\infty t^{z-1} e^{-bt} \mathrm{d}t . \label{32}
\end{equation}
Applying the Poisson summation formula (details omitted) we can
then derive
\begin{equation}
F = \frac{1}{4 \sqrt{\pi}} V_\perp \int
\frac{\mathrm{d}^2k_\perp}{(2\pi)^2} \sum_{M_N,k_x}
\sum_{p=-\infty}^\infty \int_0^\infty \mathrm{d}t ~t^{-s-1}
e^{-\frac{p^2}{4T^2 t} -t \left( k_x^2 + k_\perp^2 + M_N^2
\right)}. \label{33}
\end{equation}
Further manipulations lead us to the desired expression
\begin{equation}F=TV_\perp
\int\frac{d^2k_\perp}{(2\pi)^2}\sum_{M_N, k_x}\ln \left[2\sinh
\left(\frac{1}{2T}\sqrt{M_N^2+k_x^2+k_\perp^2}\right)\right].
\label{34}
\end{equation}
One may note here that a boson with energy $E_p$ contributes with
$(\beta=1/T)$
\begin{equation}
Z_p= \sum_{n=1}^\infty e^{-\beta E_p(n+1/2)} = \frac{1}{2 \sinh(\beta E_p /2)} \label{35}
\end{equation}
to the total partition function \cite{brevik01}. Summing over all
energies we obtain the classical expression corresponding to
Eq.~(\ref{34}). The expression of the free energy of a scalar bulk field is equal
to that of bosons with energy $E_p^2=M_N^2+k_x^2+k_\perp^2$, where
$M_N$ is the masses found in Section \ref{Section21}.
For $M_N=0$ this is free energy of a scalar field in
Minkowski (i.e. flat) spacetime without extra spatial dimensions.
By letting $T\to 0$ we find the zero-point energy
\begin{equation}
E= V_\perp \int\frac{d^2k_\perp}{(2\pi)^2}\sum_{M_N, k_x} \frac{1}{2} \sqrt{M_N^2+k_x^2+k_\perp^2}.
\end{equation}
Again, we observe that $M_N=0$-term in the sum correspond to the familiar expression
for 3+1-dimensional Minkowski spacetime.

Another expression for $F$ can be derived which in our context is
more useful, in view of our application of the Abel-Plana
summation formula later on. We start from the expression
(\ref{31}), introduce a generalized polar coordinate
transformation along the same lines as in Ref.~\cite{li97}, and
integrate over all angles. We then obtain  (the limit $s\rightarrow 0$
is understood)
\begin{equation}
F=-\frac{TV_\perp}{4\pi}\frac{\partial}{\partial
s}\mu^{2s}\sum_{l=-\infty}^\infty \sum_{M_N,k_x}\int_0^\infty dr
~r(C+r^2)^{-s}, \label{36}
\end{equation}
where $C$ is defined as
\begin{equation}
C=M_N^2+\epsilon_l^2+k_x^2. \label{37}
\end{equation}
The integral is solved using the variable change $x=r^2/C$ and
leads essentially to the Beta function
$B(q,v)=\Gamma(q)\Gamma(v)/\Gamma(v+q)$. We obtain
\begin{equation}
F=-\frac{TV_\perp}{8\pi}\Gamma(-1)\sum_{l=-\infty}^\infty
\sum_{M_N,k_x}(M_N^2+\epsilon_l^2+k_x^2). \label{38}
\end{equation}
This is the finite-temperature form that we will use below. The
corresponding zero-temperature form is found by a limiting
procedure to be
\begin{equation}
E=-\frac{V_\perp}{16\pi^\frac{3}{2}} \Gamma \left(-\frac{3}{2} \right)
\sum_{M_N,k_x}(M_N^2+k_x^2)^\frac{3}{2}.
\label{39}
\end{equation}
From now on we will set $V_\perp=1$. Hence $E$, $F$ and $P$ (force) refer to
respectively energy, free energy and force per
unit area of the physical plates.

\subsection{The piston model}
\begin{figure}[!ht]
    \centering
    \includegraphics[width=0.75\linewidth]{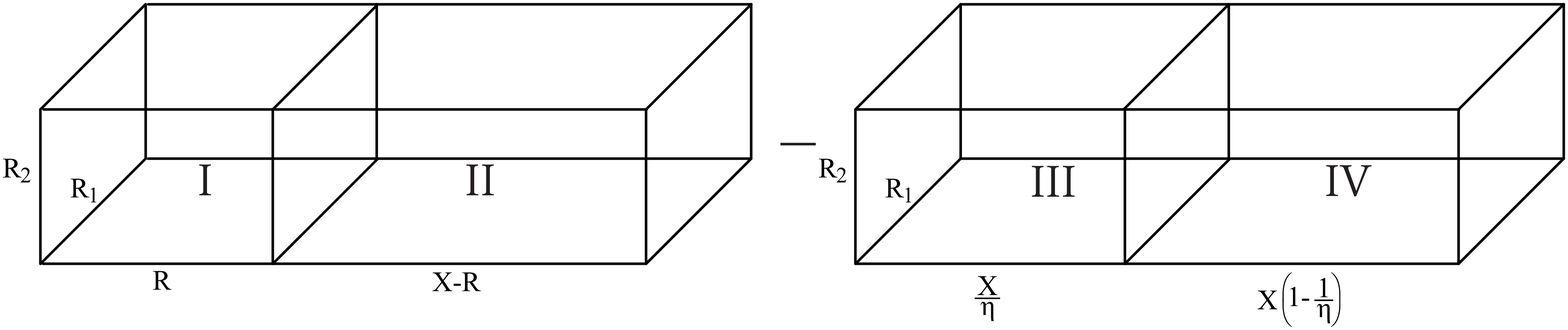}
    \caption{Illustration of the four cavities in the piston model.}
    \label{fig:piston}
\end{figure}

Before finding explicit expressions and specifying BCs we
introduce the piston model. The model has attained a great deal of
attention
\cite{cavalcanti04,cheng08,fulling09,hertzberg07,lim09,teo09b}. We
introduce the piston (Figure \ref{fig:piston}) with the same
notation as in Chapter 4.3 of \cite{elizalde94}.
Instead of only using the free energy $F_\mathrm{I}$
of cavity $I$ as the Casimir free energy, we use
\begin{equation}
F_\mathrm{piston} = F^I(a)+F^{II}(X-a)-F^{III}(X/\eta)-F^{IV}(X(1-1/\eta)).
\label{40}
\end{equation}
Initially the system is in an unstressed situation
where the cavities have size $X/\eta$ and $X(1-1/\eta)$.
Then we shift the middle plate so that the lengths
of the two cavities are $a$ and $X-a$; the system
is now in a stressed situation.
The Casimir free energy is the sum the free energies
of two cavities in the stressed case (I and II)
minus the free energies of the cavities in the unstressed
case (III and IV). The constant $\eta$ is $\sim 2$,
characterizing the unstressed situation. In the end we
let $X\to \infty$ and effectively remove
the rightmost plate from the setup. In the piston model
all terms independent or linear in $a$ vanish, hence from now
on we will discard all such terms.

\section{Casimir free energy and force: Initial remarks}

\subsection{The Abel-Plana formula}

We want to find a more explicit expression for the Casimir free
energy, one we can evaluate numerically. Thus all the summations
over $k_x$ and $M_N$ from Eq.~(\ref{38}) need to be taken care of.
Instead of using Eq.~(\ref{38}) we look at the complex function
\begin{equation}
F(s)=-\frac{T}{8\pi}\Gamma(s)\sum_{l=-\infty}^\infty
 \sum_{M_N,k_x}(M_N^2+\epsilon_l^2+k_x^2)^{-s}, \label{41}
\end{equation}
which reduces to the free energy in Eq.~(\ref{38}) when $s=-1$. The function
$F(s)$ is well-defined for large, positive $\mathrm{Re}(s)$ and we
analytically continue it to the whole complex plane. Together with
$F(s)$ we will use a variant of the Abel-Plana formula \cite{saharian07,elizalde09}
\begin{equation}
\begin{split}
&\sum_{n=1}^\infty \frac{\pi f(z_n)}{1+(1/z_n) \sin {z_n}
\cos(z_n+2\alpha)}
= \underbrace{-\frac{\pi}{2} \frac{f(0)}{1-\beta_0/a-\beta_a/a} }_1\\
&~+ \underbrace{\int_0^\infty \mathrm{d}z f(z) }_2
+ \underbrace{ i\int_0^\infty \mathrm{d}z \frac{f\left( e^{i\pi/2} z\right)-f\left( e^{-i\pi/2} z\right)}
{\frac{(\beta_0/a-1)(\beta_a/a-1)}{(\beta_0/a+1)(\beta_a/a+1)}e^{2z}-1} }_3,
\label{42}
\end{split}
\end{equation}
especially suited for plates with Robin BC. Here, $z_n$ denotes
the $n$'th zero in the right half of the complex plane of the
complex function $F_x(z=ak_x)$ in Eq.~(\ref{19}). From
Eqs.~(\ref{18}) and (\ref{19}) we can find the relation
\begin{equation}
1+(1/z_n)\sin{z_n} \cos(z_n+2\alpha) =
1-\frac{\beta_0/a}{1+(\beta_0 z_n/a
)^2}-\frac{\beta_a/a}{1+(\beta_a z_n/a )^2}. \label{43}
\end{equation}
By choosing
\begin{equation}
f(z) = \frac{1}{\pi} \left( M_N^2 + \epsilon_l^2 + z^2/a^2
\right)^{-s} \left(1-\sum_{j=0,a} \frac{\beta_j /a}{1+ (\beta_j
z/a)^2} \right), \label{44}
\end{equation}
the left hand side of Eq.~(\ref{42}) matches the sum over $k_x$ in $F(s)$.
The notation $\sum_{j=0,a}$ means there are contributions from both
the left ($j=0$) and the right ($j=a$) plate.

\subsection{Application of the Abel-Plana formula}

We can divide the free energy $F$ at arbitrary temperature $T$
into two separate parts, $F=F(M_N=0)+F(M_N>0)$. For a massive
scalar field there is no massless mode ($M_N=0$) at all. For a
massless field, even and minimally coupled, there is an $M_N=0$
mode. Recall that $F(M_N=0)$ yields the same expression as the
free energy of the massless scalar in Minkowski spacetime. To find
the Casimir energy and force for such a field one can simply add
the massless mode term. It is natural therefore to analyze the
$M_N=0$ mode separately. The formal expressions are divergent, and
will be regularized by the use of zeta functions.

Let now $M_N$ be arbitrary. Insert the expression (\ref{44}) into
(\ref{42}), and divide the sum into three separate parts as
indicated by the underlines 1,2, and 3. We do not give the details
here, as the formalism is analogous to that of
Ref.~\cite{elizalde09}, pertaining to the zero temperature case.
The free energy can be written as the sum of three parts: one part
$F_{NP}$ as the contribution when no plates are present, one part
$F_j$ as the vacuum free energy along the transverse directions
induced by the plates at $x_0=0$ and $x_a=a$ respectively, and a
remaining part $\Delta F$. Thus
\begin{equation}
F=F_{NP}+\sum_{j=0,a}F_j+\Delta F. \label{45}
\end{equation}
The two first terms do not refer to the gap width $a$, or are
linearly dependent on $a$, and do not contribute to the free
energy in the piston model. The last term $\Delta F$, henceforth
called simply $F$, is the term of physical importance. It is
precisely the term corresponding to underline 3 in Eq.~(\ref{42}).
We give this expression explicitly:
\[ F(s)=-\frac{ T}{(2\pi)^2}\Gamma(s)\sin \pi s \]
\[
\times \sum_{l=-\infty}^\infty
\sum_{M_N}\int_{a\sqrt{M_N^2+\epsilon_l^2}}^\infty \,dz\,
\frac{[z^2/a^2-(M_N^2+\epsilon_l^2)]^{-s}}{\frac{(\beta_0/a-1)(\beta_a/a-1)}{(\beta_0/a+1)(\beta_a/a+1)}e^{2z}-1}
\]
\begin{equation}
\times \left( 1-\sum_{j=0,a}\frac{\beta_j/a}{1-(\beta_jz/a)^2}
\right). \label{46}
\end{equation}
We can now use this expression as basis for discussing special
cases: DD, NN and DN boundary conditions. We first consider flat
space with no additional spatial dimensions.

\section{DD,  NN and DN boundary conditions in flat space with no extra dimensions}

To demonstrate the procedure used for finding the Casimir free
energy and force we look at the well known case: A massless,
scalar field in flat spacetime (Minkowski metric) and no extra
spatial dimensions. An additional motivation for including this
section is that $F(M_N=0) = F_{Mink}$, as mentioned earlier.

Consider first the general formalism. With  DD or NN boundary
conditions we obtain, when making use of the substitution
$z=xa\sqrt{M_N^2+\epsilon_l^2}$,
\[ F^{DD,NN}=-\frac{aT}{(2\pi)^2}\Gamma(s)\sin \pi s \]
\begin{equation}
\times \sum_{l=-\infty}^\infty
\sum_{M_N}(M_N^2+\epsilon_l^2)^{-s+1/2}\int_1^\infty
dx\frac{(x^2-1)^{-s}}{e^{2a\sqrt{M_N^2+\epsilon_l^2}}-1}.
\label{47}
\end{equation}
We expand the denominator and use the relation
\begin{equation}
\int_1^\infty (x^2-1)^{\nu-1} e^{-\mu x} \mathrm{d}x
= \frac{1}{\sqrt{\pi}} \left( \frac{2}{\mu} \right)^{\nu-\frac{1}{2}} \Gamma (\nu) K_{\nu-\frac{1}{2}} (\mu),
\label{48}
\end{equation}
In the limit $s\rightarrow -1$ we use the property $\Gamma(x)\sin
\pi x=\pi/\Gamma(1-x)$ to get the free energy for arbitrary $T$
\begin{equation}
F^{DD,NN}=-\frac{\sqrt{\pi}aT}{(2\pi)^2}\sum_{l=-\infty}^\infty
\sum_{M_N}\sum_{n=1}^\infty
\left(\frac{M_N^2+\epsilon_l^2}{n^2a^2}\right)^{\frac{3}{4}}K_{\frac{3}{2}}\left(2na\sqrt{M_N^2+\epsilon_l^2}\right).
\label{49}
\end{equation}

 The same expression follows if one makes use of zeta
regularization. The Abel-Plana formula is powerful in the present
context, as it is easily adjustable to different choices for the
boundary conditions.

Consider next flat space. With  $M_N=0$ we obtain from
 Eq.~(\ref{49})
\begin{equation}
F^{DD,NN}_{Mink}=-\frac{\zeta_R(3)T}{16\pi
a^2}-\frac{1}{2}aT\left(\frac{2T}{a}\right)^{\frac{3}{2}}\sum_{l,n=1}^\infty
\left(\frac{l}{n}\right)^{\frac{3}{2}}K_{\frac{3}{2}}(4\pi anlT).
\label{50}
\end{equation}
The first term corresponds to $l=0$, and is derivable for instance
by taking into account the properties of $K_\nu(z)$ for small
arguments.

For high temperatures, $aT \gg 1$, the expression (\ref{50}) is
suitable. The first term is the dominant one, as the $K_\nu$ terms
decrease for increasing temperatures.

For low temperatures, $aT \ll 1$, some rewriting is however
necessary. We go back to the complex function
\begin{equation}
F(s)=-\frac{T}{8\pi}\Gamma(s)\sum_{l=-\infty}^\infty
\sum_{k_x}(\epsilon_l^2+k_x^2)^{-s}, \label{51}
\end{equation}
which corresponds to Eq.~(\ref{38}) when $M_N=0, s=-1$. Splitting
off the $l=0$ term and using again the Mellin transform (\ref{32})
we can write $F(s)$ as
\begin{equation}
F(s)=F_{l=0}-\frac{T}{4\pi}
\sum_{k_x}\int_0^\infty dt \,t^{s-1}S_2(4\pi^2T^2t) e^{-k_x^2 t}. \label{52}
\end{equation}
Here
\begin{equation}
F_{l=0}=-\frac{T}{8\pi}\Gamma(s)\sum_{k_x}(k_x^2)^{-s}, \label{53}
\end{equation}
and $S_2(t)$ is the function
\begin{equation}
S_2(t)=\sum_{m=1}^\infty e^{-m^2t}, \label{54}
\end{equation}
possessing the property \cite{elizalde94}
\begin{equation}
S_2(t)=-\frac{1}{2}+\frac{1}{2}\sqrt{\frac{\pi}{t}}+
\sqrt{\frac{\pi}{t}}S_2\left(\frac{\pi^2}{t}\right). \label{55}
\end{equation}
The first of the three terms coming from the rhs of Eq.~(\ref{55})
cancels $F_{l=0}$, leaving
\[
F(s)=-\frac{1}{16\pi^{3/2}}\Gamma(s-\frac{1}{2})\sum_{k_x}k_x^{-2(s-1/2)}
\]
\begin{equation}
-\frac{1}{8\pi^{3/2}}\sum_{k_x}\sum_{l=1}^\infty
\int_0^\infty dt\,t^{(s-1/2)-1}\exp
\left(-tk_x^2-\frac{l^2}{4T^2t}\right). \label{56}
\end{equation}
The first term here is recognized as the zero-temperature energy,
$F(T=0)=E$. With $s=-1$ we find
\begin{equation}
F_{Mink}=E_{Mink}-\frac{1}{4\pi^{\frac{3}{2}}} (2T)^{\frac{3}{2}}
\sum_{k_x}\sum_{l=1}^\infty
\left(\frac{k_x}{l}\right)^{\frac{3}{2}}K_{\frac{3}{2}}\left(\frac{k_xl}{T}\right).
\label{57}
\end{equation}
We can now make use of the Abel-Plana formula (\ref{42}), choosing
for the function $f(z)$ the form
\begin{equation}
f(z)=\frac{1}{\pi}\left(\frac{z}{a}\right)^{\frac{3}{2}}K_{\frac{3}{2}}\left(\frac{lz}{aT}\right)\left(1-\sum_{j=0,a}\frac{\beta_j/a}{1+(\beta_jz/a)^2}\right).
\label{58}
\end{equation}
This leads to, when omitting terms not contributing to the piston
model,
\begin{equation}
F_{Mink}=E_{Mink}-\frac{2a}{\pi^2}\sum_{n,l=1}^\infty
\frac{1}{(4a^2n^2+l^2/T^2)^2}. \label{59}
\end{equation}
We once more use the Mellin transform, but this time choosing
$S_2(4a^2t)$ together with Eq.~(\ref{55}). Some calculation leads
to the final expression
\begin{equation}
F^{DD,NN}_{Mink}=E^{DD,NN}_{Mink}-\frac{2T^{\frac{3}{2}}}{(2a)^{\frac{3}{2}}}
\sum_{n,l=1}^\infty
\left(\frac{n}{l}\right)^{\frac{3}{2}}K_{\frac{3}{2}}\left(\frac{\pi
ln}{aT}\right), \label{60}
\end{equation}
where $E^{DD,NN}_{Mink}=-\pi^2/(1440a^3)$. The corresponding expression for the pressure is
\begin{equation}
\begin{split}
P^{DD,NN}_{Mink}=&P^{DD,NN}_{Mink}(T=0) -\frac{3
T^\frac{3}{2}}{\sqrt{2}\, a^\frac{5}{2}} \sum_{n,l=1}^\infty
\left( \frac{n}{l} \right)^\frac{3}{2}
K_\frac{3}{2} \left( \frac{\pi l n}{aT} \right)\\
&+\frac{ \pi \sqrt{T/2}}{ a^\frac{7}{2}} \sum_{n,l=1}^\infty
\frac{n^\frac{5}{2}}{\sqrt{l}}\, K_\frac{5}{2} \left( \frac{\pi l
n}{aT} \right), \label{61}
\end{split}
\end{equation}
where $P^{DD,NN}_{Mink}(T=0)=-\pi^2/(480a^4)$. The Casimir energy and force are equal to the zero temperature
expressions plus correction terms, the latter decaying
exponentially as $T\rightarrow 0$.

In Eq.~(\ref{61}) we may insert  the asymptotic expansion for
large arguments, $K_{\nu}(z) =
(\pi/2z)^{1/2}\,e^{-z}[1+(4\nu^2-1)/8z]$. It is of interest to
extract the dominant term in the correction, corresponding to
$n=l=1$. Approximately we then get
\begin{equation}
P_{Mink}^{DD,NN} = P_{Mink}^{DD,NN}(T=0)+\frac{\pi}{2a^3}\exp
\left(-\frac{\pi}{aT}\right). \label{61a}
\end{equation}
The physically important point here is that the finite temperature
term  is {\it positive}, corresponding to a repulsive force
correction (recall that we are considering $aT \ll 1$). The
situation is in some sense similar to that encountered in earlier
studies when calculating the Casimir force between two parallel
metallic slabs in physical space, assuming the Drude dispersion
relation for the material: also in that case the finite
temperature effect was found to {\it weaken} the attractive $T=0$
force \cite{brevik06}.

We now consider the third class of BC's mentioned above: assuming
Dirichlet boundary conditions on one plate and Neumann on the
other we find
\[ F^{DN}=\frac{aT}{(2\pi)^2}\Gamma(s)\sin \pi s \]
\begin{equation}
\times \sum_{l=-\infty}^\infty
\sum_{M_N}(M_N^2+\epsilon_l^2)^{-s+1/2}\int_1^\infty
dx\frac{(x^2-1)^{-s}}{e^{2a\sqrt{M_N^2+\epsilon_l^2}}+1}.
\label{62}
\end{equation}
The steps are similar to those of the DD and NN calculations, only
with a factor $(-1)^n$ due to the positive sign in the
denominator and accordingly $E^{DN}_{Mink}=-7/8 E^{DD,NN}_{Mink}$.
The free energy density with DN boundary conditions
becomes
\begin{equation}
F^{DN}=-\frac{a\sqrt{\pi}T}{(2\pi)^2}\sum_{l=-\infty}^\infty
\sum_{M_N}\sum_{n=1}^\infty (-1)^n
\left(\frac{M_N^2+\epsilon_l^2}{n^2a^2}\right)^{\frac{3}{4}}K_{\frac{3}{2}} \left( 2n
a\sqrt{M_N^2+\epsilon_l^2}\right) . \label{63}
\end{equation}
With $M_N=0$ we get
\begin{equation}
F^{DN}_{Mink}=\frac{3\zeta_R(3)T}{64\pi
a^2}-\frac{1}{2}aT\left(\frac{2T}{a}\right)^{\frac{3}{2}}\sum_{l,n=1}^\infty
(-1)^n\left(\frac{l}{n}\right)^{\frac{3}{2}}K_{\frac{3}{2}}(4\pi
anlT), \label{64}
\end{equation}
which is a convenient form for the case of high temperatures.

For low temperatures we obtain by a similar reasoning as that
given above,
\begin{equation}
F^{DN}_{Mink}=E^{DN}_{Mink}
-\frac{2T^\frac{3}{2}}{(2a)^\frac{3}{2}} \sum_{n,l=1}^\infty
(-1)^n\left( \frac{n}{l} \right)^\frac{3}{2} K_\frac{3}{2} \left(
\frac{\pi l n}{aT} \right) \label{65}
\end{equation}
with corresponding force
\begin{equation}
\begin{split}
P^{DN}_{Mink}=&P^{DN}_{Mink}(T=0)
-\frac{3 T^\frac{3}{2}}{\sqrt{2}\, a^\frac{5}{2}} \sum_{n,l=1}^\infty (-1)^n\left( \frac{n}{l} \right)^\frac{3}{2} K_\frac{3}{2} \left( \frac{\pi l n}{aT} \right)\\
&+\frac{ \pi \sqrt{T/2}}{ a^\frac{7}{2}} \sum_{n,l=1}^\infty
(-1)^n \frac{n^\frac{5}{2}}{\sqrt{l}} \, K_\frac{5}{2} \left(
\frac{\pi l n}{aT} \right). \label{66}
\end{split}
\end{equation}
Again extracting the dominant term by including only $n=l=1$ we
get approximately
\begin{equation}
P_{Mink}^{DN} = P_{Mink}^{DN}(T=0)-\frac{\pi T}{2a^3}\exp \left(
-\frac{\pi}{aT}\right). \label{66a}
\end{equation}
The correction term is the same as in Eq.~(\ref{61a}), but with
the opposite sign. The thermal correction is attractive.

\section{DD, NN and DN boundary conditions in RSI}

Consider first the high-temperature regime. Whereas in flat space
this corresponds to $aT \gg 1$,  in RSI the natural choice for
high temperatures is $T \gg ke^{-k \pi r_c}$. Recall that the
lowest values of $M_N$ are $\sim ke^{-k \pi r_c}$; this implies
$aT \gg 1$ since $ake^{-k \pi r_c} \gg 1$ for all relevant
distances in physical space. In this limit Eq.~(\ref{49}) is a
suitable expression for the free energy and the Casimir force is
\begin{equation}
\begin{split}
&P_\mathrm{RSI}^\mathrm{DD,NN} =\frac{{\sqrt{\pi}}\,T}{(2\pi)^2}
\sum_{l=-\infty}^{\infty} \sum_{M_N} \sum_{n=1}^{\infty}
\left( \frac{M_N^2 +(2\pi T l)^2}{n^2 a^2} \right)^\frac{3}{4} K_\frac{3}{2} \left(2na \sqrt{M_N^2 + (2\pi T l)^2} \right)\\
&-\frac{2{\sqrt{\pi}}\,T}{(2\pi)^2} \sum_{l=-\infty}^{\infty}
\sum_{M_N} \sum_{n=1}^{\infty} \frac{\left( M_N^2 +(2\pi T
l)^2\right)^\frac{5}{4}}{ \sqrt{na} } K_\frac{5}{2} \left(2na
\sqrt{M_N^2 + (2\pi T l)^2} \right). \label{67}
\end{split}
\end{equation}
After some rewriting we find this in accordance with Eq.~(23) in \cite{teo09a}.
We need only to include the $E(M_N=0)$ term to
get the Casimir force for a massless scalar instead of a massive.

We find that the high temperature limit is valid for $T \gg
10^{16} K$. Only temperatures much less than these are expected to
be of physical importance. It is most natural therefore to find
the Casimir energy and force for $T \ll  ke^{-k \pi r_c}$. Note
that the brane low-temperature condition does not fix the
magnitude of the product $aT$ relative to unity. With $k\sim
10^{19}$ GeV, $e^{-k\pi r_c}\sim 10^{-16}$ we only get the weak
condition $T\ll 10^3$ GeV. As an example, choose $T=300$ K
($2.6\times 10^{-11}$ GeV), $a=1 \mu$m, from which it follows that
$aT=0.15.$ In most cases of practical interest we will have $aT
\ll 1$, although one can easily consider cases where $aT \gg 1$,
still compatible with the condition $T\ll ke^{-k\pi r_c}$.

Using the same procedure as in flat space we find the RSI
equivalent to Eq.~(\ref{57}),
\begin{equation}
F_{RSI} = E_{RSI} - \frac{1}{4 \pi^\frac{3}{2}}
(2T)^\frac{3}{2}\sum_{M_N} \sum_{k_x} \sum_{l=1}^\infty \left(
\frac{k_x^2+ M_N^2}{l^2} \right)^\frac{3}{4} K_\frac{3}{2} \left(
\frac{l}{T} \sqrt{ k_x^2 + M_N^2} \right). \label{68}
\end{equation}
We can differentiate this expression to find the Casimir force. By
assuming $\partial k_x/ \partial a = -k_x/a$ we get Eq.~(17) in
\cite{teo09b}, only missing the first term. The assumption holds
for all $k_x$ proportional to $1/a$, which is the case for DD, NN
and DN BC. However, we are using the piston model, and a term like
the first one of Eq.~(17) in \cite{teo09b} should not occur in the
Casimir force. The force term in question is independent of $a$,
and thus corresponds to a free energy term that is linear in $a$.
As a consequence of using the piston model, all terms independent
or linear in $a$ must be removed from the free energy. Although
the metric in \cite{teo09b} does not include the warp factor
$e^{-2kr_c \phi}$ the expressions are the same, since the warp
factor only affects the values of the $M_N$s. In Eq.~(\ref{68})
$E_{RSI}$ is the zero temperature energy in RSI and is found from
Eq.~(\ref{39}) using the Abel-Plana formula (\ref{42}) with
\begin{equation}
f(z) = \frac{1}{\pi} \left( M_N^2 + z^2/a^2 \right)^\frac{3}{2}
\left(1-\sum_{j=0,a} \frac{\beta_j /a}{1+ (\beta_j z/a)^2} \right).
\label{69}
\end{equation}
After some variable changes the energy reads
\begin{equation}
\begin{split}
E_{RSI}=- &\frac{1}{6\pi^2} \sum_{M_N} \int_{a M_N}^\infty
\mathrm{d}z~(z^2/a^2-M_N^2)^\frac{3}{2} \frac{1
-\sum_{j=0,a}\frac{1-\beta_j/a}{(\beta_j
z/a)^2}}{\frac{(\beta_0/a-1)(\beta_a/a-1)}{(\beta_0/a+1)(\beta_a/a+1)}e^{2z}-1}.
\label{70}
\end{split}
\end{equation}
and for DD and NN boundary conditions it simplifies to
\begin{equation}
E^\mathrm{DD,NN}_{RSI} = - \frac{1}{8\pi^2 a} \sum_{n=1}^\infty
\sum_{M_N} \frac{M_N^2}{n^2} K_2(2aM_N n). \label{71}
\end{equation}
The Casimir force at zero temperature is
\begin{equation}
\begin{split}
P^\mathrm{DD,NN}_{RSI}(T=0) =& - \frac{3}{8 \pi^2 a^2}
\sum_{n=1}^\infty \sum_{M_N}
\frac{M_N^2}{n^2} K_2\left(2a M_N n\right) \\
&- \frac{1}{4 \pi^2 a} \sum_{n=1} \sum_{M_N}
\frac{ M_N^3}{n} K_1\left(2a  M_N  n\right).
\label{72}
\end{split}
\end{equation}
This is in accordance with \cite{elizalde09} through that paper
does not consider the Casimir effect rising from a bulk scalar in
the RS model in particular. Inserting the approximation
Eq.~(\ref{28}) for $M_N$ we see that the energy is essentially the
same as in \cite{turan07}. There are three minor differences.
First of all the energy in \cite{turan07} has some extra terms
linear and independent of $a$ since the piston model is not used.
Secondly,  factors $p$ are included to make the expression hold
for electromagnetic fields, where $p$ is the polarizations of the
photon. The last difference is a factor of 2 included to account
for 'the volume of the orbifold'. Since we can not see how this
factor occurs, it is not included. This is also equal to Eq.~(26)
in \cite{teo09a};  only it contains the $E(M_N=0)$ term since a
massless field is considered.

The summation over $k_x$ in Eq.~(\ref{68}) is still left and can be done using
the Abel-Plana formula with
\begin{equation}
f(z) = \frac{1}{\pi} \left( M_N^2 + z^2/a^2 \right)^\frac{3}{4}
K_\frac{3}{2} \left( \frac{l}{T} \sqrt{ z^2/a^2 + M_N^2} \right)
\left(1-\sum_{j=0,a} \frac{\beta_j /a}{1+ (\beta_j z/a)^2} \right).
\label{73}
\end{equation}
After inserting $K_\frac{3}{2}(z) = (\pi/2z)^{1/2}\, e^{-z} ( 1 +
1/z)$ the free energy reads
\begin{equation}
\begin{split}
F_{RSI} = E_{RSI} - & \frac{T^3}{\pi^2}
\sum_{M_N}\sum_{l=1}^\infty l^{-3} \int_{M_N a}^\infty \mathrm{d}z
\frac{1- \sum_{j=0,a} \frac{\beta_j/a}{1-(\beta_j z /a)^2}}{
\frac{(\beta_0/a-1)(\beta_a/a-1)}{(\beta_0/a+1)(\beta_a/a+1)}
e^{2z}-1}
\Bigg[ \sin\left( \frac{l}{T} \sqrt{ (z/a)^2 - M_N^2} \right) \\
& ~- \frac{l}{T} \sqrt{ (z/a)^2 - M_N^2} \cos\left( \frac{l}{T} \sqrt{ (z/a)^2 - M_N^2} \right) \Bigg]
\label{74}
\end{split}
\end{equation}
We continue by inserting the $\beta$'s for DD and NN boundary conditions, expansion of the
denominator and the variable exchange $x= z/a$ to get
\begin{equation}
\begin{split}
F_{RSI}^\mathrm{DD,NN} = E_{RSI}^\mathrm{DD,NN} - & \frac{T^3a}{\pi^2}
\sum_{M_N}\sum_{l=1}^\infty l^{-3} \sum_{n=1}^\infty
\int_{M_N}^\infty \mathrm{d}x e^{-2nax}
\Bigg[ \sin\left( \frac{l}{T} \sqrt{ x^2 - M_N^2} \right) \\
&- \frac{l}{T} \sqrt{ x^2 - M_N^2} \cos\left( \frac{l}{T} \sqrt{ x^2 - M_N^2} \right) \Bigg].
\label{75}
\end{split}
\end{equation}
Integrals of this form is solved in Appendix A with the result
\begin{equation}
\begin{split}
&\int_C^\infty \mathrm{d}x \left( \sin \left( A \sqrt{x^2-C^2} \right)
-A \sqrt{x^2-C^2} \cos \left( A \sqrt{x^2-C^2} \right) \right) e^{-Bx} \\
&= \frac{C^2 A^3}{A^2 + B^2} K_2\left(C\sqrt{A^2+B^2}\right)
\label{76}
\end{split}
\end{equation}
giving the free energy
\begin{equation}
\begin{split}
F^\mathrm{DD,NN}_{RSI} =&E^\mathrm{DD,NN}_{RSI}
-\frac{a}{\pi^2} \sum_{M_N} \sum_{l=1}^\infty \sum_{n=1}^\infty \frac{M_N^2}{(2na)^2 + (l/T)^2} \\
&\quad \times K_2 \left( M_N \sqrt{ (2na)^2 + (l/T)^2} \right)
\label{77}
\end{split}
\end{equation}
and force
\begin{equation}
\begin{split}
&P_\mathrm{RSI}^\mathrm{DD,NN}=P_\mathrm{RSI}^\mathrm{DD,NN}(T=0) \\
&-\frac{1}{\pi^2} \sum_{M_N} \sum_{l=1}^\infty \sum_{n=1}^\infty
\frac{M_N^2 \left(3(2na)^2-(l/T)^2\right)}{\left((2na)^2 + (l/T)^2\right)^2}K_2 \left( M_N \sqrt{ (2na)^2 + (l/T)^2} \right) \\
&-\frac{1}{\pi^2} \sum_{M_N} \sum_{l=1}^\infty \sum_{n=1}^\infty
\frac{M_N^3(2na)^2}{\left( (2na)^2 + (l/T)^2 \right)^\frac{3}{2}} K_1 \left( M_N \sqrt{ (2na)^2 + (l/T)^2} \right).
\label{78}
\end{split}
\end{equation}
This expression, belonging to the low-temperature regime  $T \ll
ke^{-k \pi r_c}$, can be used both  for $aT \ll 1$ and for $aT \gg
1$. The argument of the Bessel functions will always be large
since $ake^{-k \pi r_c} \gg 1$ for all relevant distances. The
correction terms to the zero temperature energy and force
expressions are small. The expression has to our knowledge not
been given before.
The leading term for the force in terms of $T/M_N$ is
\begin{equation}
\begin{split}
&P_\mathrm{RSI}^\mathrm{DD,NN} \sim P_\mathrm{RSI}^\mathrm{DD,NN}(T=0) \\
&-\frac{1}{\sqrt{2\pi^3}} \sum_{M_N} \sum_{l=1}^\infty \sum_{n=1}^\infty
M_N^4 e^{-\frac{M_N}{T} \sqrt{(2anT)^2+l^2} }
\left( \frac{T}{M_N} \right)^\frac{3}{2} \frac{(2naT)^2}{\left( (2naT)^2+l^2\right)^\frac{7}{4}}.
\end{split}
\end{equation}
In contrast to flat space, {\it both} zero temperature Casimir
force and the thermal correction is negative.
Hence the Casimir effect in RSI is stronger in the low temperature
limit ($T \ll ke^{-k\pi r_c}$) both for $aT \ll 1$ and $aT \gg 1$.

The DN  expressions deviate from the DD and NN  expressions in RSI
in the same way as in flat space. The factor of $(-1)^n$ must be
included in sum over $n$, where the sum over $n$ originates from
the expansion of the denominator in Eq.~(\ref{70}) and (\ref{74}).

\subsection{Comparison to flat space}
The reason for calculation the Casimir force in RSI is to
find out where there are deviations from the Casimir force
in flat spacetime without extra spatial dimensions.
For an easier comparison we give the full expression
for the Casimir force of a {\it massless bulk scalar} in RSI with
DD/NN BCs.
\begin{equation}
\begin{split}
&P_\mathrm{RSI}^\mathrm{DD,NN}=-\frac{\pi^2}{480 a^4} \\
&-\frac{1}{a^4} \sum_{M_N} \sum_{n=1}^\infty \left[ \frac{3}{8 \pi^2}
\frac{(aM_N)^2}{n^2} K_2\left(2a M_N n\right)
+ \frac{1}{4 \pi^2}
\frac{ (aM_N)^3}{n} K_1\left(2a  M_N  n\right) \right] \\
&+\frac{1}{a^4}\sum_{n,l=1}^\infty \left[ -\frac{3(aT)^\frac{3}{2}}{\sqrt{2}}
\left( \frac{n}{l} \right)^\frac{3}{2}
K_\frac{3}{2} \left( \frac{\pi l n}{aT} \right)
+ \pi \sqrt{Ta/2}
\frac{n^\frac{5}{2}}{\sqrt{l}}\, K_\frac{5}{2} \left( \frac{\pi l
n}{aT} \right) \right] \\
&-\frac{1}{a^4\pi^2} \sum_{M_N} \sum_{n,l=1}^\infty \Bigg[
\frac{(aM_N)^3(2n)^2}{\left( (2n)^2 + (l/aT)^2 \right)^\frac{3}{2}} K_1 \left( a M_N \sqrt{ (2n)^2 + (l/Ta)^2} \right) \\
&-\frac{(aM_N)^2 \left(3(2n)^2-(l/aT)^2\right)}{\left((2n)^2 + (l/aT)^2\right)^2}K_2 \left(a M_N \sqrt{ (2n)^2 + (l/aT)^2} \right) \Bigg]. \label{78a}
\end{split}
\end{equation}
This expression is good at low temperatures because the Bessel
function decreases exponentially at high arguments. Hence, we only
need to sum over the first couple of values from $M_N$, $n$ and
$l$ if the other factors ($M_N$, $a$ and $T$) ensure that the
argument is  much greater than one. On the other hand, if the
argument of the Bessel function is not large we need to be careful
that the sum has converged. Now the essential question is: For
what values (of $M_N$, $a$ and $T$) are the sums of Bessel
functions are of the same magnitude as the flat spacetime at zero
temperature (i.e. the first term)? Or simply: When can we see a
deviation from the ordinary Casimir force?

Looking at Eq.~(\ref{78a}) we see that at zero temperature we need
$a M_N \sim 1$ for noticeable difference. We know that $M_N
\sim k e^{-k\pi r_c}$ for low $N$, and $k $ is usually set to
$\sim M_{Pl} \approx 10^{19}$GeV in RSI. In the original paper of
Randall and Sundrum they propose to choose $kr_c \sim 10$ in order
to solve the hierarchy problem. With these values we find that $a$
is $\sim 10^{-21}$m. There is no point in looking at distances
smaller than the size of an atom. Only distances of physical
relevance ($>1$nm) are of interest.
\begin{figure}[!ht]
    \centering
    \subfloat[]{\label{fig:PRSI}\includegraphics[width=0.5\textwidth]{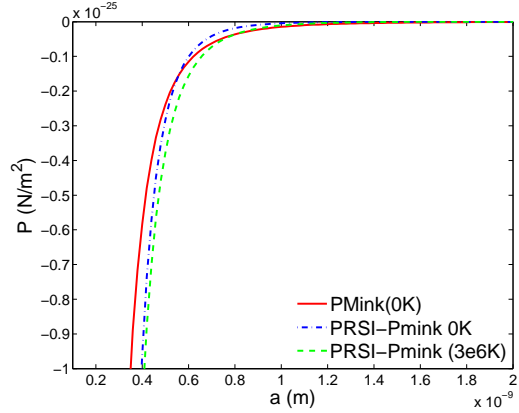}}
    \subfloat[]{\label{fig:PRSIMink}\includegraphics[width=0.5\textwidth]{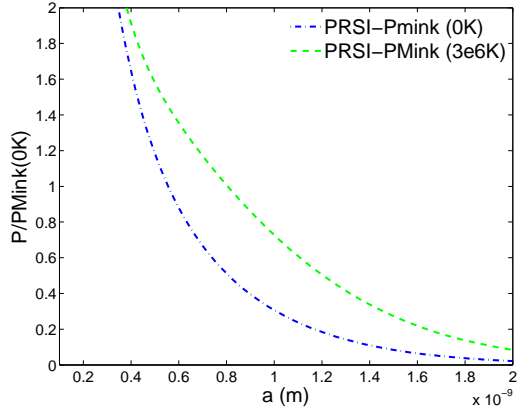}}
\caption{$P_\mathrm{RSI}^\mathrm{DD,NN}-P_{Mink}^\mathrm{DD,NN}$
for a massless scalar bulk field with $k=10^{19}$GeV and $d= e^{-k
\pi r_c}=10^{-26}$.
    Left panel: The difference between the Casimir force for RSI and flat 3+1-spacetime at $T=0$  and $T= 3\times 10^{6}~$K. The
    Casimir force for a massless scalar field in flat spacetime at zero temperature is included for comparison. Right panel: The ratio of
     the difference between the Casimir force in RSI and flat spacetime
 at $T=0$  and $T= 3\times 10^{6}~$K to the force at zero temperature in flat spacetime.}

    \label{fig:RSI}
\end{figure}
In Figure \ref{fig:RSI} we keep $k=10^{19}$GeV, but choose
$e^{-k\pi r_c}=10^{-26}$. The difference from RSI to ordinary
Casimir force $F_{Mink}$ at the zero temperature is given in Fig.
\ref{fig:RSI}. By choise of parameters the magnitude of the
correction in RSI is of the same order of magnitude as $F_{Mink}$,
given by the red line in  Fig. \ref{fig:RSI} (a). With smaller
value of $k r_c$ we will not se any difference at separations
larger than 1nm. The corresponding size of the extra dimension is
$r_c\approx 10^{-35}$m. In  Fig. \ref{fig:RSI} (b) we se the ratio
of this difference to the Casimir force at zero temperature. We
observe that the extra term we get in RSI, goes faster to zero than
$F_{Mink}$. Now we turn to relevant values of the temperature. The
choices of $r_c$, $k$ and $a$ are still the same, and after some
testing it turns out that $a_{max}T\sim 3$ is suitable. To be sure
that the sums have converged we let both $n$ and $l$ run to 30.
The result is presented in Fig. \ref{fig:RSI}. The green line is
the $P_{RSI}-P_{Mink}$ for $T=3 \times  10^{6}$K and we see that
this gives a stronger Casimir force than at zero temperature.

\subsection{Comparison to flat space with one extra dimension}
In the previous section we compared the Casimir force in RSI
with the Casimir force for a massless scalar i ordinary flat
3+1-spacetime. In this section we will look at a higher dimensional
spacetime which is flat, i.e. no warp factor in the metric.
As mentioned in the introduction, this topic has gained a lot
of interest lately. The extra dimension is a torus with
circumfence $2\pi L$. In this case $M_N=N/L$, with $N=0,\pm 1, \pm 2, \ldots$.
With these values for the $M_N$s instead of the ones we have in RSI
we see that Eq.~(\ref{67}) is equal to the high temperature expression
in \cite{teo09}. However we have not found an expression corresponding to Casimir force in Eq.~(\ref{78a}). Hence, we will derive such an expression using the Chowla-Selberg formula in \cite{lim07}. With the new values of the $M_N$s, the function $F(s)$
(for DD BCs) reads
\begin{equation}
F(s)=-\frac{T}{8\pi}\Gamma(s)\sum_{l=-\infty}^\infty
 \sum_{N=-\infty}^\infty \sum_{n=1}^\infty ((2\pi T)^2+(n\pi / a)^2+(N/L)^2)^{-s}. \label{78b}
\end{equation}
Rewriting this to homogenous Epstein zeta functions
\begin{equation}
Z_{E,p}(s;a_1,\ldots ,a_p)=  \left. \sum_{k_1, \ldots ,k_p=-\infty }^\infty \right.^\prime ((a_1 k_1)^2+(a_2 k_2)^2 + \ldots + (a_p k_p)^2 )^{-s} \label{78c},
\end{equation}
we find
\begin{equation}
F(s)=-\frac{T}{16\pi}\Gamma(s) Z_{E,2}(s;2\pi T,1/L) -\frac{T}{16\pi}\Gamma(s) Z_{E,3}(s;2\pi T, \pi /a,1/L). \label{78d}
\end{equation}
The notation $\sum_{k_1, k_2, \ldots k_p=-\infty}^\infty$ means that for $k_1$ to $k_p$ we
sum from $-\infty$ to $\infty$. The prime $\prime$ behind the sum means
that the term $k_1=k_2=\ldots=k_p=0$ is omitted.
The Chowla-Selberg formula is
\begin{equation}
\begin{split}
& Z_{E,p}(s;a_1,\ldots ,a_p) = Z_{E,m}(s;a_1,\ldots ,a_m) \\
& + \frac{\pi^m \Gamma \left(s-\frac{m}{2} \right) }{ \left( \prod_{i=1}^m a_i \right) \Gamma (s) }
Z_{E,p-m}\left( s-\frac{m}{2}; a_{m+1},\ldots , a_p \right)\\
& + \frac{2 \pi^s}{\Gamma (s) \left( \prod_{i=1}^m a_i \right) } \left. \sum_{k_1,\ldots , k_m=-\infty }^\infty \right.^\prime \left. \sum_{k_{m+1},\ldots , k_p=-\infty }^\infty \right.^\prime \left( \frac{\sum_{i=1}^m k_i/a_i }{\sum_{j=m+1}^p k_j a_j }  \right)^\frac{2s-m}{4} \\
&\quad \times K_{s-m/2}\left( 2\pi \sqrt{\sum_{i=1}^m k_i/a_i} \sqrt{\sum_{j=m+1}^p k_j a_j} \right). \label{78e}
\end{split}
\end{equation}
After use of this with $m=2$, we put $s=-1$ and remove all terms linear and independent of $a$. Then we see that the free energy is
\begin{equation}
\begin{split}
F =& - \frac{T}{16 \pi} \Gamma(-1) Z_{E,2}(-1;2 \pi T, \pi /a) \\
&- \frac{a}{4 \pi ^2}\left. \sum_{N=-\infty}^\infty \right.^\prime \left. \sum_{n,l=-\infty}^\infty \right.^\prime
\frac{(N/L)^2 }{(2na)^2 + (l/T)^2 } K_2 \left( N/L \sqrt{ (2na)^2 + (l/T)^2 } \right). \label{78f}
\end{split}
\end{equation}
The first term can be identified as $F(M_N=0)$ and the second term is equal to
Eq.~(\ref{77}) (when $M_N=N/L$). From $\left. \sum_{n,l=-\infty}^\infty \right.^\prime $ we get the factor 4 when we let the sums over $n$ and $l$ run from 1 to $\infty$. The term $l=0$, but $n \neq, 0$ gives the zero temperature expression and $n=0$ with $l \neq 0$ is independent of $a$ and should be removed in the piston model.

Since Eq.~(\ref{77}) leads to Eq.~(\ref{78a}) we can conclude that we get the same
answer with the Chowla-Selberg formula as with the Abel-Plana formula in flat spacetime
with one extra spatial dimension.
However, the Abel-Plana formula can be used regardless of the values of $M_N$, while
the Chowla-Selberg formula is only useful when we can rewrite our expressions to homogenous
Epstein zeta functions. The second advantage of the Abel-Plana formula is
that different boundary condititions can easily be obtained.

\section{DD,  NN and DN boundary conditions in RSII}
In RSII the Kaluza-Klein modes are continuous and we must replace
the sum over $M_N$ with an integral,
\begin{equation}
\sum_{M_N} \to \int_0^\infty \frac{\mathrm{d}M}{k}.
\label{79}
\end{equation}
In Eq.~(\ref{49}) we get an integral on the from
\begin{equation}
\int_0^\infty \mathrm{d} z \left( z^2 + A^2 \right)^\frac{3}{4} K_\frac{3}{2} \left( B \sqrt{z^2 + A^2} \right)
= \sqrt{\frac{\pi}{2B}} A^2 K_2(AB).
\label{80}
\end{equation}
The derivation of this formula is given in Appendix A. We find
that the free energy and force in RSII are
\begin{equation}
F^\mathrm{DD,NN}_{RSII} = -\frac{T \pi^3}{1440 k a^3} -\frac{\pi
T^3}{a k} \sum_{l=1}^\infty \sum_{n=1}^\infty \frac{l^2}{n^2}
K_2\left(4\pi T a l n \right) \label{81}
\end{equation}
and
\begin{equation}
\begin{split}
P^\mathrm{DD,NN}_{RSII} =& -\frac{3 T \pi^3}{1440 k a^4} - \frac{3
\pi T^3}{a^2 k}
\sum_{l=1}^\infty \sum_{n=1}^\infty  \frac{l^2}{n^2} K_2\left( 4\pi a T l n \right) \\
&-\frac{4 \pi^2 T^4}{a k}
\sum_{l=1}^\infty \sum_{n=1}^\infty  \frac{l^3}{n} K_1\left( 4\pi a T l n \right).
\label{82}
\end{split}
\end{equation}
These expressions are convenient for the high temperature limit
$aT \gg 1$. In RSII there are only two temperature regimes, $aT
\gg 1$ and $aT \ll 1$, since $M_N$ is continuous. To find the low
temperature limit we insert Eq.~(\ref{79}) into Eq.~(\ref{77}) and
use the integral
\begin{equation}
\int_0^\infty \mathrm{d} z ~z^2 K_2(Az) = \frac{3\pi}{2A^3}.
\label{83}
\end{equation}
The free energy reads
\begin{equation}
F^\mathrm{DD,NN}_{RSII} = E^\mathrm{DD,NN}_{RSII} - \frac{3
a}{2 \pi k} \sum_{n=1}^\infty \sum_{l=1}^\infty \frac{1}{\left(
(2an)^2+(l/T)^2 \right)^\frac{5}{2}}, \label{84}
\end{equation}
where $E^\mathrm{DD,NN}_{RSII}=-\frac{3 \zeta_R(5)}{128 \pi k a^4}$ is the zero temperature
energy in RSII. We can find this energy
from e.g. Eq.~(\ref{71}) by making use of Eq.~(\ref{83}).
We use the Mellin transform as in low temperature, flat spacetime
(with $S_2(4a^2t)$) and find
\begin{equation}
F^\mathrm{DD,NN}_{RSII} = E^\mathrm{DD,NN}_{RSII} - \frac{\pi
T^2}{2 k a^2 } \sum_{n,l=1}^\infty \left( \frac{n}{l} \right)^2
K_2 \left( \frac{\pi n l}{aT} \right). \label{85}
\end{equation}
The Casimir force is
\begin{equation}
P^\mathrm{DD,NN}_{RSII} = P^\mathrm{DD,NN}_{RSII}(T=0) +
\frac{\pi^2 T}{2 k a^4 } \sum_{n,l=1}^\infty \frac{n^3}{l} K_1
\left( \frac{\pi n l}{aT} \right), \label{86}
\end{equation}
with $P^\mathrm{DD,NN}_{RSII}(T=0)=-\frac{3 \zeta_R(5)}{32 \pi k a^5}$.

In the low temperature limit $aT \ll 1$ we get from Eq.~(\ref{86})
the dominant term corresponding to $n=l=1$,
\begin{equation}
P^\mathrm{DD,NN}_{RSII} = P^\mathrm{DD,NN}_{RSII}(T=0) +
\frac{\pi^2 T}{2 k a^4 }\sqrt{\frac{aT}{2}} \exp \left(
-\frac{\pi}{aT}\right). \label{86a}
\end{equation}
The temperature correction term is repulsive.

As in RSI the only difference between DD/NN BC and DN is a factor
$(-1)^n$ in the sum over $n$.

\section{Concluding remarks}

Our main objective has been to calculate the finite temperature
Casimir effect for a scalar field residing in the bulk in the two
Randall-Sundrum models, RSI and RSII. Two parallel plates are
envisaged, with gap $a$,  located on one of the  RS branes. We
have given  most attention  to the RSI model. Robin boundary
conditions, cf. Eqs.~(\ref{13}) and (\ref{14}), are assumed on the
two plates. The geometrical picture is the piston model, as
illustrated in Fig.~1. We have made use of the Abel-plana
summation formula throughout, as this turns out to be the most
convenient choice in the present context.

In the case of flat space  the basic expressions for Casimir free
energy and force (per unit surface area) are worked out in the
form of series in Sect.~4, both for high and for low temperatures.
A characteristic feature for DD and NN boundary conditions on the
two plates is that the dominant part of the finite temperature
correction term for low temperatures ($aT \ll 1$) is repulsive.
That is, the force becomes decreased slightly when the temperature
increases from zero. In this sense the behavior is analogous to
that encountered in the case of conventional Casimir theory for
metallic slabs in physical space when the dispersive relation for
the material is taken to have the Drude form \cite{brevik06}.

The RSI model is covered in Section 5 in an analogous way.
The dominant term in the Casimir force shows the characteristic property
to strengthen the zero temperature effect instead of weaken it
as in RSII. From Eq.(\ref{78}) we can evaluate the Casimir
force both for $aT\ll 1$ and $aT \gg 1$, provided $T\ll ke^{-k\pi r_c}$.
This section also covers a comparison to flat space with and
without a compactified extra dimension.

In Section 6 the RSII model is considered. We have $\int_0^\infty
\mathrm{d}M/k = \pi/k \int_{-\infty}^\infty \mathrm{d}M/(2\pi)$
and thus the Casimir force has a characteristic $\pi/k$ times the
Casimir force of a 4+1-dimensional Minkowski spacetime. This is
pointed out by Morales-T\'{e}cotl {\it et al.} \cite{morales08},
but as an argument against using zeta functions in the
regularization and rather use Green's functions. It is not only
the regularization method that is different, the physical picture
is also differs from this article and the work by Frank {\it et
al.} \cite{turan07,saad08}. While Frank {\it et al.} calculate the
free energy of a slice of the bulk, Morales-T\'{e}cotl {\it et
al.} try to restrict the system to the brane by evaluating the
Green's function at the visible brane ($y=0$ for RSII). In this
way they claim to incorporate the localization properties of the
modes of the scalar bulk field ($\psi_N(y)$). Note that the
Green's function method includes a integral over $y$ and by
setting $y=0$, Morales-T\'{e}cotl {\it et al.} thus remove the
$y$-dependence of a part of the integrand before integrating.
However, keeping the $y$-dependence before the integration makes
one obtain results different from those of Frank {\it et al.}. The
delicate point is how to include the localization properties of
the modes. In Ref.~\cite{flachi09} it is proposed  to resolve the
issue by changing the boundary conditions. The problem with the
localization properties of the modes needs to be resolved before
the Casimir effect form an electromagnetic field can be
considered.

\newpage
\renewcommand{\theequation}{\mbox{\Alph{section}.\arabic{equation}}}
\appendix
\setcounter{equation}{0}

\section{Integrals and Bessel functions}
We need to calculate the nontrivial integral
\begin{equation}
\int_C^\infty \mathrm{d}x \left(A \sqrt{x^2-C^2} \cos \left( A
\sqrt{x^2-C^2} \right) -\sin \left( A \sqrt{x^2-C^2} \right)
\right)e^{-Bx}.
\end{equation}
First  use the substitution $u=\sqrt{x^2-C^2}$ to find
\begin{equation}
\begin{split}
&\int_0^\infty \frac{\mathrm{d}u}{\sqrt{u^2+C^2}} (A u^2 \cos(Au) - u \sin(Au) ) e^{-B\sqrt{u^2+C^2}} \\
=& A \frac{\partial^2}{\partial B^2} \int_0^\infty \mathrm{d}u ~ \frac{1}{\sqrt{u^2+C^2}} \cos(Au) e^{-B\sqrt{u^2+C^2}} \\
-&A \int_0^\infty \mathrm{d}u ~ \frac{1}{\sqrt{u^2+C^2}} \cos(Au) e^{-B\sqrt{u^2+C^2}} \\
-&\int_0^\infty \frac{\mathrm{d}u}{\sqrt{u^2+C^2}} u \sin(Au) )
e^{-B\sqrt{u^2+C^2}}.
\end{split}
\end{equation}
Using Eq.~3.961 in \cite{gradshteyn80}
\begin{equation}
\int_0^\infty \frac{x \mathrm{d}x}{\sqrt{\gamma^2+x^2}} e^{-\beta
\sqrt{\gamma^2 + x^2}} \sin(ax) = \frac{a \gamma}{\sqrt{a^2+
\beta^2}} K_1\left( \gamma \sqrt{a^2 + \beta^2} \right)
\end{equation}
as well as
\begin{equation}
\int_0^\infty \frac{\mathrm{d}x}{\sqrt{\gamma^2+x^2}} e^{-\beta
\sqrt{\gamma^2 + x^2}} \cos(ax) = K_0\left( \gamma \sqrt{a^2 +
\beta^2} \right),
\end{equation}
we get
\begin{equation}
\begin{split}
&\int_C^\infty \mathrm{d}x \left(A \sqrt{x^2-C^2} \cos \left( A
\sqrt{x^2-C^2} \right)
-\sin \left( A \sqrt{x^2-C^2} \right) \right)e^{-Bx} \\
=& A \frac{\partial^2}{\partial B^2} K_0
\left(C\sqrt{A^2+B^2}\right)
-\frac{AM}{\sqrt{A^2+B^2}} K_1\left(C\sqrt{A^2+B^2}\right) \\
&-AM^2 K_0 \left(C\sqrt{A^2+B^2}\right).
\end{split}
\end{equation}
After differentiating and using the relationships between $K_0$,
$K_1$ and $K_2$ we find the solution
\begin{equation}
\begin{split}
&\int_C^\infty \mathrm{d}x \left(A \sqrt{x^2-C^2} \cos \left( A
\sqrt{x^2-C^2} \right)
-\sin \left( A \sqrt{x^2-C^2} \right) \right) e^{-Bx} \\
=& -\frac{C^2 A^3}{A^2 + B^2} K_2\left(C\sqrt{A^2+B^2}\right).
\end{split}
\end{equation}

Next consider the integral
\begin{equation}
\int_0^\infty \mathrm{d} z \left( z^2 + A^2 \right)^\frac{3}{4}
K_\frac{3}{2} \left( B \sqrt{z^2 + A^2} \right).
\end{equation}
With the substitution  $u = \sqrt{ (z/A)^2 + 1}$ we get
\begin{equation}
\begin{split}
& A^\frac{5}{2} \int_1^\infty \frac{ \mathrm{d} u ~u^\frac{5}{2} }{\sqrt{u^2-1}} K_\frac{3}{2}(BAu) \\
=& \sqrt{\frac{\pi}{2B}} A^2 \int_1^\infty \frac{ \mathrm{d} u
~u^2 }{\sqrt{u^2-1}} \left(1 + \frac{1}{BAu} \right) e^{-BAu} ,
\end{split}
\end{equation}
again using $K_\frac{3}{2}(z) = (\pi/2z)^{1/2}\, e^{-z} (1+1/z)$.
As the integral representation of $K_2(BA)$ is
\cite{Abramowitz72})
\begin{equation}
K_2(ax) = \frac{ \pi^\frac{1}{2} \left(\frac{1}{2} ax
\right)^2}{\Gamma \left( \frac{5}{2} \right) } \int_1^\infty
\mathrm{d} u ~e^{-axu} (u^2-1)^\frac{3}{2},
\end{equation}

we find after several partial integrations
\begin{equation}
K_2(ax) = \frac{ \pi^\frac{1}{2} \left(\frac{1}{2} ax
\right)^2}{\Gamma \left( \frac{5}{2} \right) } \frac{3}{(ax)^2}
\int_1^\infty \frac{ \mathrm{d} u ~u^2 }{\sqrt{u^2-1}} \left(1 +
\frac{1}{axu} \right) e^{-axu}.
\end{equation}
Thus the integral becomes
\begin{equation}
\int_0^\infty \mathrm{d} z \left( z^2 + A^2 \right)^\frac{3}{4}
K_\frac{3}{2} \left( B \sqrt{z^2 + A^2} \right) =
\sqrt{\frac{\pi}{2B}} A^2 K_2(BA).
\end{equation}

\newpage

\appendix

\end{document}